\documentclass[aps, prl, floatfix, twocolumn, showpacs]{revtex4-1}

\usepackage{graphicx}
\usepackage{color}
\usepackage{calc}
\usepackage{array}
\usepackage{graphicx}
\usepackage{amsmath, amssymb}
\usepackage{natbib}
\usepackage{hyperref}

\def\abs#1{\left \vert #1 \right \vert}

\def\sup#1{\textsuperscript{#1}}

\begin{document}

\title{Rotons in interacting ultracold Bose gases}

\author{Samuel C. \surname{Cormack}}
\author{D{\'{a}}niel \surname{Schumayer}} \email{dschumayer@physics.otago.ac.nz}
\author{David A. W. \surname{Hutchinson}}
\affiliation{Jack Dodd Centre for Quantum Technology,
             Department of Physics,
             University of Otago,
             Dunedin, New Zealand}
\begin{abstract}
   In three dimensions, non-interacting bosons undergo Bose-Einstein
   condensation at a critical temperature, $T_{c}$, which is slightly
   shifted by $\Delta T_{\mathrm{c}}$, if the particles interact. We
   calculate the excitation spectrum of interacting Bose-systems,
   \sup{4}He and \sup{87}Rb, and show that a roton minimum emerges
   in the spectrum above a threshold value of the gas parameter. We
   provide a general theoretical argument for why the roton minimum
   and the maximal upward critical temperature shift are related. We
   also suggest two experimental avenues to observe rotons in
   condensates. These results, based upon a Path-Integral Monte-Carlo
   approach, provide a microscopic explanation of the shift in the
   critical temperature and also show that a roton minimum does
   emerge in the excitation spectrum of particles with a
   structureless, short-range, two-body interaction.
\end{abstract}

\date{\today}
%

\maketitle

Appearance of a roton minimum \cite{Feynman1998} in the excitation
spectrum is an essential part of the phenomenology of superfluidity.
Bogoliubov derived \cite{Bogoliubov1947} the excitation spectrum of
a weakly interacting boson system for small momenta
\begin{equation} 
   \label{eq:Bogoliubov}
   \varepsilon(k)
   =
   \sqrt{
         \displaystyle
         \frac{g n}{m} k^{2}
         +
         \varepsilon_{0}(k)^{2}
         \,
        }
   \hspace*{3mm}
   \xrightarrow{k \rightarrow 0}
   \hspace*{3mm}
   \sqrt{\frac{g n}{m}} \,k\:,
\end{equation}
where $\varepsilon_{0}(k)$ denotes the free-particle energy, $n=N/V$
is the particle density, $g$ measures the interaction strength, and
$m$ is the mass of a particle. The two-body interaction is approximated
by a repulsive $\delta$-potential with strength determined by the
$s$-wave scattering length, $a$.

The importance of interactions on the macroscopic scale is characterised
by the gas parameter, $\gamma = n a^{3}$, which is proportional to the
ratio of the volume occupied by the particles compared to that available
to them. Bose-Einstein condensates in ultra-cold gases are usually in the
weakly interacting limit \cite{Cornell2002}; and away from the Feschbach
resonance experiments could only explore the dilute limit, $\gamma \sim
10^{-6}$ \cite{Anderson1995}. However the use of Feshbach resonances in
\sup{85}Rb has facilitated the creation of condensates with tunable values
of $\gamma$ up to approximately $8 \times 10^{-3}$ \cite{Papp2008}. In
condensates of \sup{7}Li, values of $\gamma$ up to $\sim 50$ have been
achieved \cite{Pollack2009}, although with non-uniform density distribution
due to trapping. We note here that $\gamma$ is not at all small, $\sim 0.2$,
in the case of liquid \sup{4}He \cite{Reppy2000, Leggett2001}.

Interactions also induce a shift in the condensation critical
temperature, $T_{\mathrm{c}}$, which was first reported for the
\sup{4}He-Vycor system \cite{Reppy2000} confirming earlier theoretical
predictions \cite{Stoof1992, Bijlsma1996, Gruter1997, Holzmann1999,
Arnold2001, Kashurnikov2001}
\begin{equation}
   \label{eq:TemperatureShift}
   \Delta T_{\mathrm{c}}
   \equiv \frac{T_{\mathrm{c}} - T_{\mathrm{c}}^{(0)}}{T_{\mathrm{c}}^{(0)}}
   \cong C \,\gamma^{1/3}
\end{equation}
where $T_{\mathrm{c}}^{(0)}$ is the critical temperature of an ideal Bose-gas
with the same density and $C$ is a positive dimensionless constant.
Different theoretical approaches aimed at determining this dimensionless
constant resulted in significant discrepancy for some time, with
consensus finally provided by Monte Carlo simulations
\cite{Kashurnikov2001}.

For small $\gamma$, the Bogoliubov spectrum \eqref{eq:Bogoliubov}
accurately describes the excitation spectrum, while for larger $\gamma$,
such as in liquid \sup{4}He, it deteriorates and a roton minimum is
observed \cite{Henshaw1961}. Path integral Monte Carlo calculations
\cite{Gruter1997, Pilati2008} and an experiment \cite{Reppy2000} have
shown that equation \eqref{eq:TemperatureShift} becomes invalid for
$\gamma \gtrsim 10^{-3}$. For even higher $\gamma$, the system
freezes and the roton minimum goes soft.

\begin{figure}[b!]
   \includegraphics[angle=-90, width=89mm]{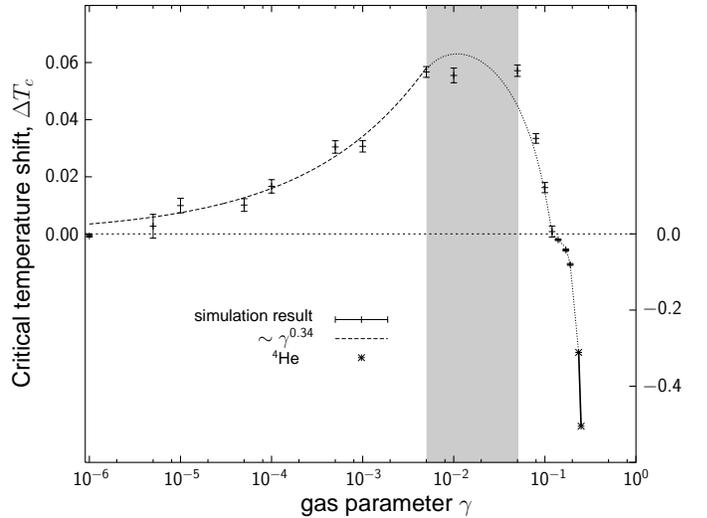}
   \caption{\label{fig:CriticalTemperatureShift}%
            The critical temperature shift $\Delta T_{\mathrm{c}}$,
            is plotted as a function of $\gamma$ as derived by
            Gr{\"u}ter {\textit{et al}}. \cite{Gruter1997}. Note
            the different scales on the ordinate. Reproduced through
            the courtesy of the authors of Ref.~\cite{Gruter1997}.
            The shaded area covers the range where the maximum
            shift is located.
           }
\end{figure}

In fact, $\Delta T_{\mathrm{c}}$, has a maximum when $\gamma \approx10^{-2}$ as
can be seen in Fig.~\ref{fig:CriticalTemperatureShift}.
Since the onset of condensation is associated with the energies of
available states, we might postulate the peak in $\Delta T_{\mathrm{c}}$ to
be accompanied by the transition to the roton regime as $\gamma$
increases. Below we substantiate this claim.

Contrary to earlier studies in which the condensate was rotating fast
\cite{Sinha2005} or particles had laser-induced dipole moments
\cite{ODell2003}, we show that the roton minimum may also occur in a
homogeneous system where the two-body interaction is short-range and
structureless. This fact may give hope for indirectly observe rotons
in Bose-Einstein condensates utilising already existing experimental
aparatus. Moreover, our simulations imply that the excitation of
rotons is also connected to the shift of the critical temperature of
condensation scrutinized in earlier studies.

{\textit{Theoretical model:}} We assume that our system can be
described using only two-body interaction potentials between the
constituents and can be modelled by the hard-sphere potential with
diameter $a$. We have varied $a$, while keeping $n$ constant,
to sample different values of $\gamma$. The systems were homogeneous
and periodic boundary condition have been applied.

{\textit{Numerical details:}} We have carried out path integral Monte
Carlo (PIMC) simulations \cite{Ceperley1995} for both \sup{4}He and
\sup{87}Rb in order to determine the energy spectrum, $\varepsilon(k)$.
The configurations of the system are sampled via the density matrix.
Although the exact density matrix is only available at high
temperature, the low temperature density matrix can be constructed
by convolution of a sufficient number of high temperature density
matrices. This iteration forms the basis of PIMC simulations.

In our simulations, we used 128 (\sup{4}He) or 64 (\sup{87}Rb)
particles  with 64 density matrices and ensured that increasing the
number of density matrices did not alter the final result. Number
densities and temperatures of $1.52 \times 10^{11} \mathrm{cm^{-3}}$,
10nK (\sup{4}He) and $3.85\times10^{11} \mathrm{cm^{-3}}$, 8\,nK
(\sup{87}Rb) are utilised. The density for \sup{87}Rb is at the
lower end of those reported in experiments \cite{Anderson1995,
Thompson2005}, while the \sup{4}He density is similar for comparison.
Simulations with realistic liquid \sup{4}He density have also been
carried out with similar conclusion presented below.

{\textit{Analysis:}} The essential object of our analysis is the
static pair correlation function, $g(r)$, which is proportional to
the probability of finding two atoms at a distance $r$ from each
other. PIMC methods are well suited for the calculation of $g(r)$
from which the static structure factor can also be acquired via
$S(k) = 1 + \frac{4\pi n}{k} \int{\! r (g(r)-1) \sin{\! (kr)} \,dr}$
\cite{Feenberg1969}. Finally, $S(k)$ is linked to the excitation
spectrum by the following relation \cite{Bijl1940, Feynman1954}
\begin{equation}
   S(k)
   =
   \displaystyle
   \frac{\hbar^{2} k^{2}}{2m \,\varepsilon(k)}
   \coth{\!\left ( \frac{\varepsilon(k)}{2kT} \right )}
   \hspace*{3mm}
   \xrightarrow{T \rightarrow 0}
   \hspace*{3mm}
   \displaystyle
   \frac{\hbar^{2} k^{2}}{2m \, \varepsilon(k)}
\end{equation}
providing an upper bound on $\varepsilon(k)$.

Due to the finite size of the system, $g(r)$ can only be calculated
over a limited range (repeated use of the periodic boundary
condition would lead to unphysical spatial correlation). One may
assume that $g(r > r_{\mathrm{c}})=1$ for some cut-off $r_{\mathrm{c}}$,
however, this approach introduces a non-continuous step in $g(r)$ at
$r_{\mathrm{c}}$ and leads to unacceptable error in $S(k)$ below
$\sim r_{\mathrm{c}}^{-1}$. The truncation error can be suppressed
by varying the cut-off, $r_{\mathrm{c}}$, and taking the average
of the values of $S(k)$ obtained this way.

\begin{figure}[bt]
	\includegraphics[angle=-90, width=\textwidth/2-2mm]%
	                {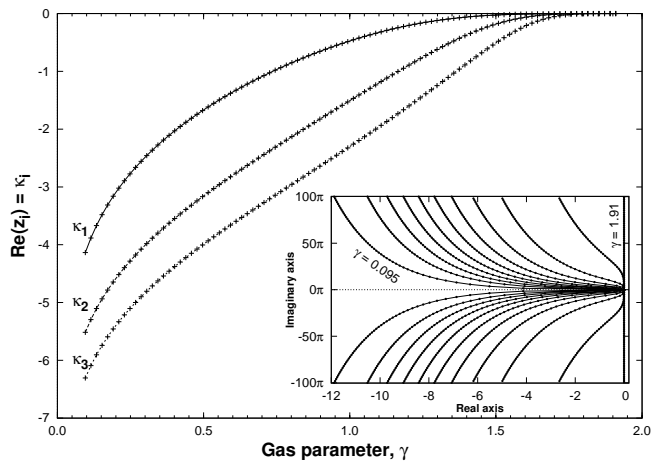}
	\caption{\label{fig:ZerosOfThePYApproximation}
	         The real parts of the first three lowest complex poles
	         are shown. The inset exhibits all the zeros over a range of the complex plane
	         for different $\gamma$. The leftmost curve belongs to
	         $\gamma = 0.095$, while the rightmost to $\gamma = 1.91$.
	         As $\gamma$ increases the zeros move towards the
	         imaginary axis and all zeros become purely imaginary for
	         $\gamma = 1.91$.
	        }
\end{figure}

An alternative is to fit a trial function to $g(r)$. The trial
function used below has been derived from the Ornstein-Zernike
equation with the Percus-Yevick (PY) closure \cite{Percus1958}
providing the analytical, but implicit solution $g_{\mathrm{PY}}
(r) = {\mathcal{L}}^{-1} \!\lbrace {\mathcal{G}}_{\mathrm{PY}}(z)
\rbrace$, where ${\mathcal{L}}^{-1}$ stands for the inverse
Laplace-transform and ${\mathcal{G}}_{\mathrm{PY}}(z)$ is known
explicitly \cite{Wertheim1963, Thiele1963}. The function,
${\mathcal{G}}_{\mathrm{PY}}(z)$, has a pole at the origin,
$z_{0} = 0$, and infinitely many distinct conjugate pairs of
poles, $z_{\ell} = \kappa_{\ell} + \mathrm{i} k_{\ell}$ ($\ell
=$1, 2, \dots). The distribution of these zeros
completely determines $g_{\mathrm{PY}} (r)$, and thus all
thermodynamical quantities of the system, for a fixed value
of $\gamma$. Figure \ref{fig:ZerosOfThePYApproximation} depicts
the real part of the three lowest lying zeros, while the inset
shows all the zeros over the complex plane in the region [-12,
0] $\times$ [$-100\pi, 100\pi$]. For moderate values of $\gamma$
only $z_{1}$ contributes significantly \cite{Nezbeda1974} and
dominates the asymptotic behaviour as
\begin{equation}
   \label{eq:FirstOrderPY_RDF}
   g_{\mathrm{PY}}(r)
   \cong
   1 + \frac{C_{1}}{r}
       \cos{\left ( k_{1} r + \delta_{1} \right )}
       e^{-\kappa_{1} r}
   .
\end{equation}
In Fig. \ref{fig:ZerosOfThePYApproximation} the real parts of the
first three lowest lying poles are shown. One may notice that
$\kappa_{1}$ approaches zero rapidly, and therefore the corresponding
oscillatory contribution to $g(r)$ is only weakly damped. The
imaginary part $k_{\ell}$ (smaller inset) determine the angular
wavenumber of the oscillatory contribution. The smallest imaginary
part, $k_{1}$, falls between $\pi$ and $3\pi$ providing a
characteristic wavenumber, i.e. a broad qualitative estimate on
where the roton minimum may occur, i.e. between $\pi/a$ and $3\pi/a$.

\begin{figure}[bht!]
   \includegraphics[angle=-90, width=\textwidth/2-2mm]{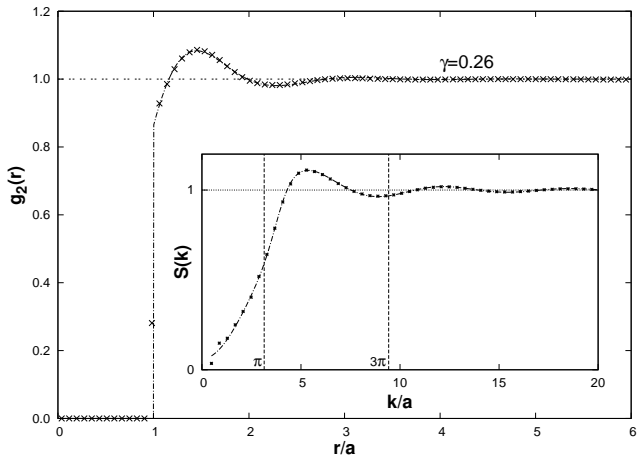}
   \caption{\label{fig:FittedCurveOnRDF}%
            The radial distribution function ($\times$) and the
            fitted first order PY approximation (dash-dotted line)
            are depicted for $\gamma = 0.26$. The statistical error
            is within the size of the crosses. Inset shows the
            corresponding static structure factor. The vertical
            dashed lines indicate the range for the first maximum
            predicted by the PY approximation.
           }
\end{figure}

In order to check our PIMC results and also to smooth the simulation
data of $g(r)$ we determine the fitting parameters $C_{1}$, $k_{1}$,
$\delta_{1}$, and $\kappa_{1}$ for each value of $\gamma$. Although
$\kappa_{1}$ and $k_{1}$ could be determined from ${\mathcal{G}}_{
\mathrm{PY}}(z)$, we treat them as free parameters, and compare their
fitted values to those the pole-structure suggests. The comparison
provides a test of the accuracy of our PIMC calculation. Satisfactory
agreement between the fitted function and raw data can be seen in Fig.
\ref{fig:FittedCurveOnRDF}. The advantage of using the fitted
$g_{\mathrm{PY}} (r)$ is that it can be extrapolated to larger system
sizes and its derivative may be easily calculated.

\begin{figure}[bt]
	\includegraphics[angle=-90, width=\textwidth/2-4mm]{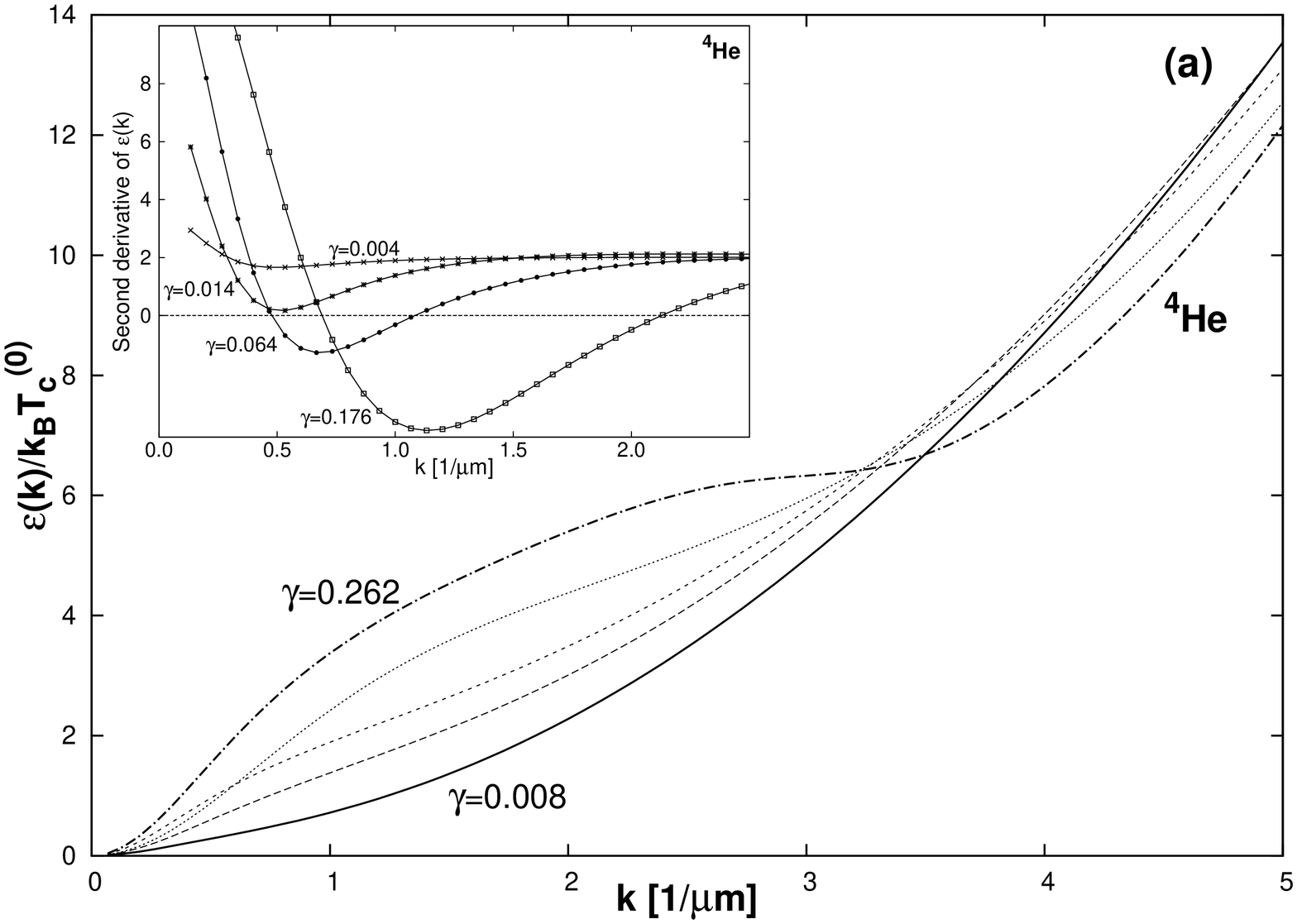}
	\includegraphics[angle=-90, width=\textwidth/2-2mm]{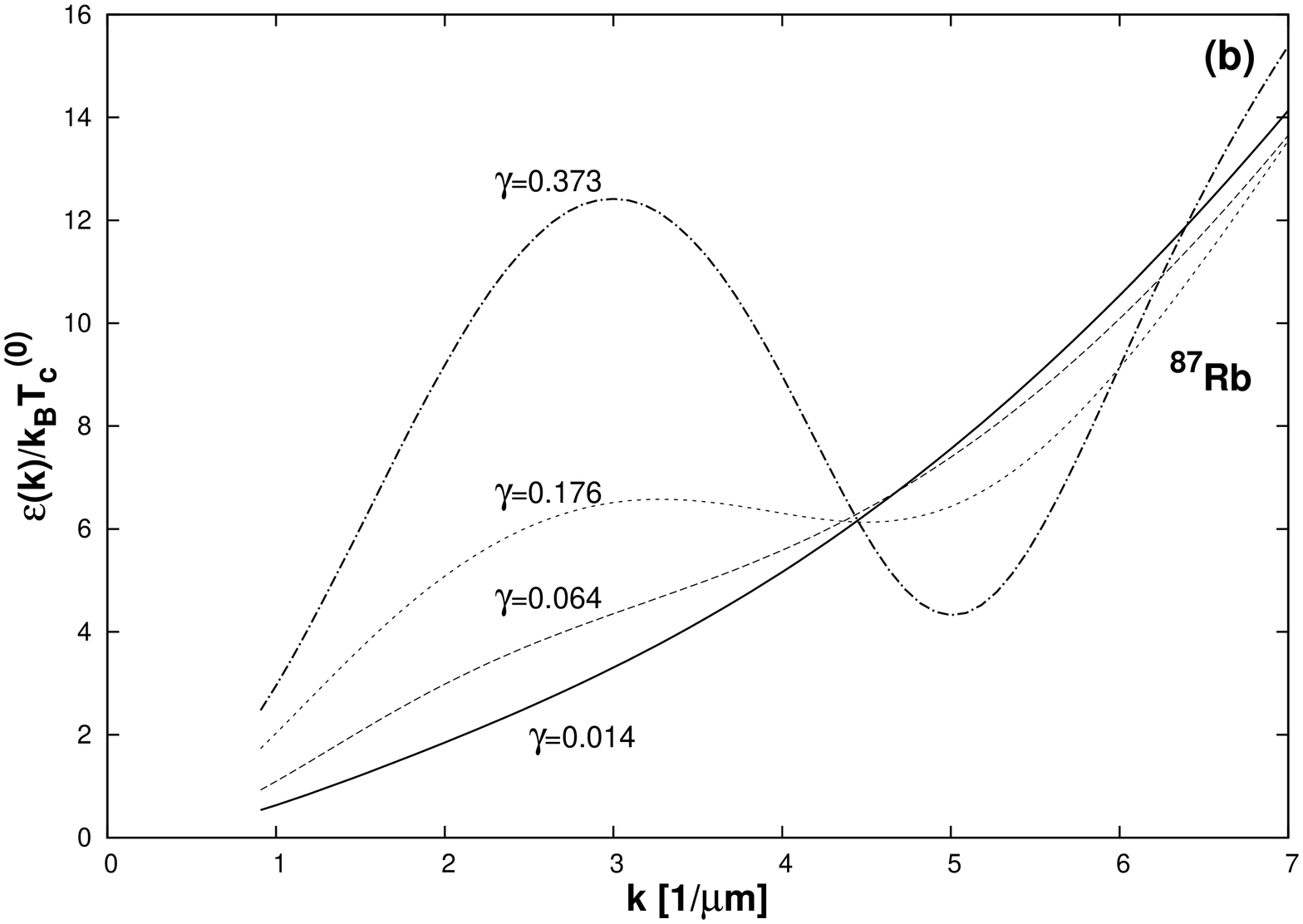}
	\caption{\label{fig:ExcitationSpectrum}%
	         The excitation spectra, $\varepsilon(k)$, are shown for
	         \textsuperscript{4}He (a) and \textsuperscript{87}Rb
	         (b). The values of gas parameter, $\gamma$, are 0.262,
	         0.176, 0.111, 0.064, 0.008 for \sup{4}He and
	         0.373, 0.176, 0.064, 0.014 for \sup{87}Rb. The inset
	         depicts the second derivative of $\varepsilon(k)$
	         for different values of $\gamma$. The ordinate is in
	         units of $\lambda=\hbar^2/2m$. Data for $\gamma =
	         0.014$ nearly reaches zero, thereby signaling the
	         emergence of an inflection point in $\varepsilon(k)$.
            }
\end{figure}

The key result of our work, the excitation spectra of \sup{4}He and
\sup{87}Rb can be seen in Fig. \ref{fig:ExcitationSpectrum}. The
maximum in $\Delta T_{\mathrm{c}}$ occurs at $\gamma \approx 0.01$, while we
find that the roton minimum appears at $\gamma \approx 0.2$. Although
the onset of the roton minimum does not appear to directly coincide
with the maximum of $\Delta T_{\mathrm{c}}$, the minimum is preceded by
another qualitative change, namely the development of a point of
inflection in $\varepsilon(k)$. By taking the second
derivative of $\varepsilon(k)$, as shown in the inset of Fig.
\ref{fig:ExcitationSpectrum}(a), we can determine the approximate
value of $\gamma$ for which the spectrum develops this inflection
point. The data indicate that this occurs approximately at $\gamma
\approx 0.014$ which falls exactly into the region where the
critical temperature reaches its maximum.

The connection between the development of the inflection point and
the maximum of $\Delta T_{\mathrm{c}}$ may be illuminated by the following
argument. At temperature $T$ the occupancy of a state with energy
$\varepsilon(k)$ is determined by the Bose-Einstein distribution,
$f(T, \varepsilon)$. Bose-Einstein condensation occurs if the ground
state is macroscopically occupied, i.e. the number of particles in
any excited states, $N_{\mathrm{ex}} (T)$, is saturated
\begin{equation}
   \label{ineq:ConditionOfBEC}
   N_{\mathrm{ex}}(T)
   =
   \int_{0}^{\infty}{G(\varepsilon) f(T, \varepsilon) \, d\varepsilon}
   \, < \,
   N_{\mathrm{total}}.
\end{equation}
It is tacitly assumed that the chemical potential has reached its
maximum value. The critical temperature, $T_{\mathrm{c}}$, is thus
determined by the excitation spectrum via the density of states,
$G(\varepsilon)$. If $G(\varepsilon)$ increases for thermally available
states, then the temperature must be decreased to reduce $f (T,
\varepsilon)$ and preserve the validity of inequality
(\ref{ineq:ConditionOfBEC}). The critical temperature should
therefore decrease as well. The density of states corresponds
to $\abs{dk/d\varepsilon}$, which is the reciprocal
of the slope of the excitation spectrum.

For weakly interacting bosons, Bogoliubov's result (\ref{eq:Bogoliubov})
indicates that increasing $\gamma$ increases the slope of
$\varepsilon(k)$, thus $G(\varepsilon)$ must decrease around
$k \approx 0$. We can therefore conclude that the critical
temperature of a weakly interacting Bose-gas $\Delta T_{\mathrm{c}}$
must be positive. As $\gamma$ increases, $\varepsilon(k)$ develops
an inflection point and becomes convex in a certain region. The
decrease of the slope in this region opens an abundance of excited
states for the particles, manifested as a peak in $G(\varepsilon)$.
These excited states can be populated at lower temperature, thus
reducing the population of the ground state. Therefore the system
has to be further cooled for macroscopical occupation of the ground
state. Consequently, the critical temperature must decrease,
resulting in $\Delta T_{\mathrm{c}}$ decreasing also.
\begin{figure}[bt]
	\includegraphics[width=\textwidth/2-1mm]{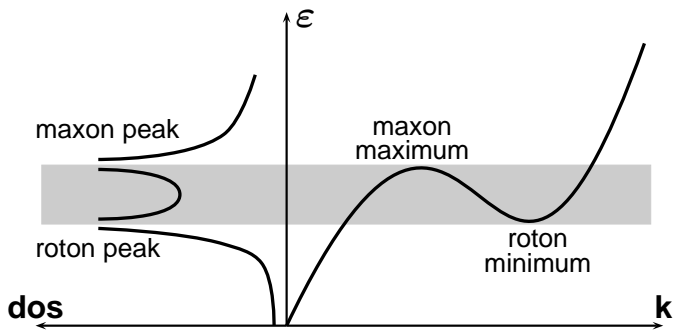}
	\caption{\label{fig:QualitativeDOS}%
	         Qualitative picture of the excitation spectrum (right)
	         and the density of states (left). As $\varepsilon(k)$ develops
	         a maximum (maxon) and a minimum (roton) the density of
	         states diverges.
            }
\end{figure}
As $\gamma$ increases further, the inflection gives rise to the roton
minimum in the spectrum for non-vanishing momenta. Although the
magnitude of the derivative starts to increase along the sides of the
minimum, the density of states continues to increase, as there are
now three sets of $k$ states which contribute to the density of states
for a given energy (See Fig. \ref{fig:QualitativeDOS}).

Recently, a similar conjecture has been made in the context of
two-dimensional dipolar systems \cite{Filinov2010}. The interaction
parameter corresponding to our $\gamma$ is the dipole coupling $D$.
As $D$ increases the critical temperature developes a maximum.
The authors mentioned the apparent coincidence of this maximum and
the onset of the roton minimum, although they have not investigated
this in detail. We note that their system and ours are dissimilar in
two important aspects: dimensionality and the nature of interaction.
Dimensionality is significant in that it influences which phases
appear in the system, e.g. Berezinksii-Kosterlitz-Thouless phase
\cite{Berezinskii1971, Berezinskii1972, Kosterlitz1973} rather than
BEC. However the nature of the interactions --we feel-- is more
fundamental. The two-body interaction in our analysis is spherically
symmetric, structureless and short-range, while their dipole
interaction has strong angular dependence and long range influence.
It is far from obvious a priori that hard core bosons, therefore,
should establish a roton structure in the excitation spectrum at
all, let alone that its appearance should correspond to the shift
of the critical temperature.

Experimental verification of the predicted onset of the roton
minimum may be possible by utilising a Feshbach resonance to tune
the interaction strength and then using Bragg spectroscopy to probe
the excitation spectrum. The values of $\gamma$ achieved by
Papp \emph{et. al.} \cite{Papp2008} approach the lower end of
values for which we predict the inflection point in $\varepsilon(k)$
to occur, while Bragg spectroscopy has previously been
used to measure the excitation spectrum of a weakly interacting
BEC \cite{Steinhauer2002}.

As a second scenario, quantum evaporation could be adapted for
pancake shaped Bose-Einstein condensates. This technique proved
to be a successful experimental method in the case of liquid
\sup{4}He \cite{Hope1984}. An excitation having enough energy is
capable of ejecting an atom from the condensate. However, phonons
and rotons carry different momenta at the same energy, therefore
the ejected atoms have different angular distribution depending
on which type of excitations they interacted with. Therefore
detecting the angular distribution of ejected atoms after exciting
the condensate in a controlled manner, could prove the existence
of rotons in BECs.

In this Letter, we have shown that the maximum observed in $\Delta
T_{\mathrm{c}}$ of an interacting BEC is related to the appearance
of an inflection point in $\varepsilon(k)$. As $\gamma$ increases
the inflection point signals the appearance of a roton minimum,
characteristic of the excitation spectrum of e.g. superfluid \sup{4}He.
We have also provided a physical argument as to why this happens
and how it could be observed experimentally.

\acknowledgments
We greatfully acknowledge the discussion with Jean Dalibard.
This work was supported under contract NERF-UOOX0703 and also
by the University of Otago.

\bibliography{pimc.bib}
\bibliographystyle{apsrev4-1}

\end{document}